\documentclass[5p]{elsarticle}

\usepackage{graphicx}
\usepackage{dcolumn}
\usepackage{bm}

\usepackage{amsmath,amssymb}
\usepackage{hyperref}
\usepackage{graphicx}
\usepackage{mathrsfs}
\usepackage{float}
\usepackage{ulem}
\usepackage{braket}
\usepackage{color}

\newcommand{\be}{\begin{equation}}
\newcommand{\ee}{\end{equation}}
\newcommand{\beq}{\begin{equation}}
\newcommand{\eeq}{\end{equation}}

\newcommand{\bea}{\begin{equation}\begin{aligned}}
\newcommand{\eea}{\end{aligned}\end{equation}}
\newcommand{\ba}{\begin{align}}
\newcommand{\ea}{\end{align}}

\begin{document}


\title{The boundary dual of the bulk symplectic form}

 \author[1]{Alexandre Belin}
\author[2]{Aitor Lewkowycz} 
\author[3]{G\'abor S\'arosi \corref{cor1}} \ead[url]{sarosi@sas.upenn.edu}

\cortext[cor1]{Corresponding author}

\address[1]{Institute for Theoretical Physics, University of Amsterdam, Science Park 904, 1098XH Amsterdam, The Netherlands,}
\address[2]{Department of Physics, Stanford University, Stanford, California, USA, }
\address[3]{David Rittenhouse Laboratory, University of Pennsylvania,   Philadelphia, PA 19104, USA,
\\Theoretische Natuurkunde, Vrije Universiteit Brussels, Pleinlaan 2, Brussels, B-1050, Belgium}

\begin{abstract}
In this paper, we study the overlaps of wavefunctionals prepared by turning on sources in the Euclidean path integral. For nearby states, these overlaps give rise to a K\"ahler structure on the space of sources, which is naturally induced by the Fubini-Study metric. The K\"ahler form obtained this way can also be thought of as a Berry curvature and, for holographic field theories, we show that  it is identical to the gravitational symplectic form in the bulk. 
We discuss some possible applications of this observation, in particular a boundary prescription to calculate the variation of the volume of a maximal slice.
\end{abstract}

\maketitle


\section{Introduction}
The Ryu-Takayanagi formula \cite{Ryu:2006bv} opened an intriguing connection between quantum gravity and quantum information. Since then, there have been many advances in understanding the holographic duals of quantum information measures in AdS/CFT, see \cite{Rangamani:2016dms} for a review. Most of these quantities are tied to subregions or mixed states and there  has not been much progress in understanding measures associated with pure states. While there have been some proposals for the duals of complexity or fidelity \cite{Susskind:2014rva,Brown:2015bva,MIyaji:2015mia}, none of these have been derived. It is therefore highly desirable to understand better the bulk duals of boundary quantities that only depend on wave functionals, hoping that they will teach us about the structure of the Hilbert space in quantum gravity.  In this paper, we are going to give a precise mapping between the antisymmetrized overlap of states (naturally understood as a K\"ahler form in the space of state deformations) and the bulk symplectic form.  This generalizes the work of \cite{MIyaji:2015mia} and identifies the precise bulk dual. The states in consideration are prepared by turning on arbitrary (possibly complex) Euclidean sources, which correspond to classical geometries. This new entry in the dictionary has similarities with the relative entropy for subregions:  it compares two neighbouring states and the bulk and boundary quantities are equal \cite{Jafferis:2015del}. We will show that our result has implications for the dual of the volume of the extremal time slice, which we will explore further in \cite{BLS}.

\section{Symplectic form from Fubini-Study metric}
Suppose we have a map from a complex manifold $\mathcal{M}$ to the Hilbert space $\mathcal{H}$. Let the manifold have complex coordinates $\alpha_i, \alpha_i^*$ and the associated state be $|\alpha\rangle$. We assume that the map is ``holomorphic" (i.e. the complex structure on the manifold and the Hilbert space is compatible) in the sense that
\beq
\label{eq:myconditions}
\partial_{\alpha_i^*} |\alpha\rangle =0, \;\;\; \partial_{\alpha_i} \langle \alpha |=0.
\eeq
Note that the $|\alpha\rangle$ states are thus necessarily un-normalized. We can pull back the Fubini-Study metric to obtain a line element on the manifold $\mathcal{M}$
\bea
ds^2 &= \frac{\langle \delta \psi | \delta \psi \rangle}{\langle \psi | \psi \rangle}-\frac{|\langle \psi | \delta \psi \rangle |^2}{\langle \psi|\psi\rangle^2}\\
&=\left( \frac{\partial_{\alpha^*_i}\langle  \alpha |\partial_{\alpha_j} |\alpha\rangle }{\langle \alpha | \alpha \rangle}- \frac{\partial_{\alpha^*_i}\langle \alpha |\alpha\rangle\langle \alpha|\partial_{\alpha_j} |\alpha\rangle }{\langle \alpha | \alpha \rangle^2}\right)d{\alpha_i}^* d\alpha_j \\
&=\partial_{\alpha_i}\partial_{\alpha_j^*}\log\langle \alpha|\alpha \rangle d\alpha_i d\alpha_j^*.
\eea
We see that the conditions \eqref{eq:myconditions} ensure that the manifold $\mathcal{M}$ endowed with the pullback of the FS metric is K\"ahler, with K\"ahler potential
\beq
\label{eq:kahlerpot}
\mathcal{K}=\log\langle \alpha|\alpha \rangle.
\eeq Therefore, there also exists a closed 2-form, the K\"ahler form on $\mathcal{M}$
\beq
\label{eq:kahler}
\Omega = i \partial_{\alpha_i}\partial_{\alpha_j^*}\log\langle \alpha|\alpha \rangle d\alpha_i \wedge d\alpha_j^*.
\eeq
Since a K\"ahler form is always a symplectic form, we deduce that $\mathcal{M}$ is symplectic. 
In the following, we denote globally the coordinates of ${\cal M}$ by $\tilde \alpha=(\alpha,\alpha^*)$ and most of the time, we will plug explicit variations into the symplectic form 
\begin{equation}
\label{eq:symplmatrix}
    \Omega(\delta_1 \tilde \alpha,\delta_2 \tilde \alpha)=i \partial_{\alpha_i}\partial_{\alpha_j^*}\log\langle \alpha|\alpha \rangle (\delta_1 \alpha_i \delta_2 \alpha_j^*-1\leftrightarrow 2).
\end{equation}

The above symplectic form can be interpreted as the Berry curvature 2-form associated to the Berry connection \cite{berry1984quantal} on the parameter space $\mathcal{M}$
\beq
\mathcal{A}=i\langle \Psi_\alpha|d|\Psi_\alpha \rangle, \quad |\Psi_\alpha\rangle = \frac{|\alpha\rangle}{\sqrt{\langle \alpha|\alpha\rangle}}.
\eeq
Indeed, in terms of the K\"ahler potential \eqref{eq:kahlerpot}
\bea
\mathcal{A} &= \frac{i}{2}\partial_{\alpha_i} \mathcal{K}d\alpha_i-\frac{i}{2}\partial_{\alpha^*_i} \mathcal{K}d\alpha^*_i=\frac{i}{2}(\partial-\partial^*)\mathcal{K},
\eea
therefore
$
\Omega = d\mathcal{A}.
$

As a warm up example, let us apply this machinery to the simple harmonic oscillator. Conditions \eqref{eq:myconditions} are satisfied by the unnormalized coherent states
\beq
|\alpha \rangle = e^{\alpha a^\dagger} |0\rangle, \;\;\;\; \langle \alpha| \alpha \rangle = e^{|\alpha|^{2}}.
\eeq
This results in the K\"ahler potential $\mathcal{K} =  |\alpha|^2$ and symplectic form
\beq
\Omega = i d\alpha \wedge d\alpha^*=-2 dp\wedge dq,
\eeq
where we have written the last line in terms of real coordinates $d\alpha = dq+i dp$.

\section{Quantum field theory}
This construction automatically gives a symplectic form on the space of sources in any QFT. Define
\bea
|\lambda\rangle &= Te^{-\int_{t_E<0} dt_E d^{d-1}\vec{x} \lambda(t_E,\vec{x}) O(t_E,\vec{x})}|0\rangle , \\ \langle \lambda | &= \langle 0|Te^{-\int_{t_E>0} dt_E d^{d-1}\vec{x} \lambda^*(-t_E,\vec{x}) O^\dagger (t_E,\vec{x})}.
\eea
Here, $t_E$ is Euclidean time and $\vec{x}$ are the spatial coordinates. These states should be thought of as being prepared by a path integral. In a CFT, the wavefunctional of the vacuum is obtained by path integrating over half of a $S^d$. The states $\ket{\lambda}$ are obtained by inserting additional sources on this manifold.   Note that these two states only depend on $\lambda(t_E),\lambda^*(t_E)$ for $t_E<0$ and we treat these two functions independently. The symbol $T$ denotes Euclidean time ordering. 
When $\lambda(t_E,\vec{x})$ is independent of $t_E$, one can think of these states as the ground states of the deformed Hamiltonian
\beq
H=H_{CFT}+\int d^{d-1}\vec{x} \lambda(\vec{x})O(0,\vec{x}).
\eeq
More generally, in holographic theories we expect states corresponding to classical geometries to be of the form $\ket{\lambda}$ and they are completely determined by the value of the sources \cite{Witten:1998qj}. 
Around any such state, we expect that for small variations of $\lambda$ (compared to the classical background), these states correspond to bulk coherent states  \cite{Skenderis:2008dg,Botta-Cantcheff:2015sav,Marolf:2017kvq}. Note that using this basis of Euclidean sources to parametrise states is quite redundant and we expect these states to have non-zero overlap. 

The $|\lambda\rangle$ states satisfy \eqref{eq:myconditions}. They give rise to the K\"ahler potential $\mathcal{K}=\log Z[\tilde \lambda]$, where $Z[\tilde \lambda]$ is the partition function with sources
\beq
\tilde \lambda(x) =
\begin{cases} \lambda(x) & t^x_E<0\\
\lambda^*(x^T) & t^x_E>0, 
\end{cases}
\label{eq:lambdatilde}
\eeq
where we use the shorthand $x^T=(-t^x_E,\vec{x})$ for coordinates reflected in Euclidean time. We think about the K\"ahler potential as a functional of half-sided sources $(\lambda,\lambda^*)$ and obtain the K\"ahler form via \eqref{eq:kahler}
\bea
\label{eq:qftkahler}
\Omega (\delta_1 \tilde \lambda,\delta_2 \tilde \lambda) &=i(\delta_1^* \delta_2-\delta_2^*\delta_1)\log Z[\tilde \lambda] \\ &=i \int_{\substack{t^x_E>0\\t^y_E<0}} dx   dy G^c_{\tilde \lambda}(x,y) \delta \lambda_{[1}^* \delta \lambda_{2]}(x^T,y),\\
\delta \lambda_{[1}^* \delta \lambda_{2]}(x^T,y)&= \delta \lambda_1^*(x^T) \delta \lambda_2(y)-\delta \lambda_1(y) \delta \lambda_2^*(x^T),
\eea
where we have defined the connected two point function 
$
G^c_{\tilde \lambda}(x,y)=\langle O^\dagger (x) O(y) \rangle_{\tilde \lambda}- \langle O^\dagger (x)  \rangle_{\tilde \lambda}\langle O(y) \rangle_{\tilde \lambda},
$
and the $\langle . \rangle_{\tilde \lambda}$ denotes (normalized) expectation value with the sources $\tilde \lambda$ turned on.
Alternatively, we may write this in terms of the change in the one point function $\delta \langle O (x)\rangle=\int dy G^c_{\tilde \lambda}(x,y) \delta \lambda(y)$ as

\beq
\label{eq:canonicalpair}
\Omega (\delta_1 \tilde \lambda,\delta_2 \tilde \lambda) = i \int_{t_E>0} dx(\delta \lambda_1^*\delta_2 \langle O \rangle-\delta \lambda_2^*\delta_1 \langle O \rangle).
\eeq

Besides the K\"ahler form, a K\"ahler manifold has a complex structure and a K\"ahler metric. The complex structure $J$ is naturally inherited from the property of the CFT inner product 
$
\langle \lambda_1 |i\lambda_2\rangle = -\langle i \lambda_1 |\lambda_2\rangle = i \langle \lambda_1 |\lambda_2\rangle,
$
and acts on variations as
\beq
\label{eq:complexstructure}
J: \;\;\delta \lambda \mapsto i \delta \lambda, \;\;\; \delta \lambda^* \mapsto -i\delta \lambda^*.
\eeq
The K\"ahler metric is given by the symmetric part of the double variation
\bea
\label{eq:kahlermetric}
g(\delta_1 \tilde \lambda,\delta_2 \tilde \lambda) &=\frac{1}{2}\Omega(\delta_1 \tilde \lambda,J(\delta_2 \tilde \lambda))\\ &=\frac{1}{2}(\delta_1 \delta_2^*+\delta_2\delta_1^*)\log Z[\tilde \lambda]. 
\eea
We note that when we only source marginal operators and take the sources to be constant, the metric \eqref{eq:kahlermetric} is proportional to the usual Zamolodchikov metric on the conformal manifold \cite{Zamolodchikov:1986gt,Seiberg:1988pf}.

\section{Holographic theories}
Here we show that \eqref{eq:qftkahler} is the bulk symplectic form for holographic theories. We assume $O$ is a single trace operator dual to a bulk field $\phi$, which can have spin but we will omit the indices for simplicity. The standard dictionary \cite{Witten:1998qj} tells us that to leading order in $N$, the overlap between states is just computed by the gravitational action in the presence of sources
\beq
\langle \lambda |\lambda \rangle \equiv Z_{CFT}[\tilde \lambda]= e^{-S^{\rm on-shell}_{\rm grav}[\tilde \lambda ]},
\eeq
where $\tilde \lambda$, defined in \eqref{eq:lambdatilde}, sets the boundary condition for the bulk fields. Therefore the K\"ahler potential is
$
\mathcal{K} = -S^{\rm on-shell}_{\rm grav}[\tilde \lambda ].
$
We now want to calculate \eqref{eq:qftkahler}. Let us study the variations of the on-shell gravitational action with respect to the boundary conditions for this. The variation of the Lagrangian density satisfies
\beq
\label{eq:symploneform}
\delta (\mathcal{L}d^{d+1} x) = -E_\phi \delta \phi d^{d+1} x +  d {\bf \theta}(\phi,\delta\phi),
\eeq
where ${\rm \bf \theta}$ is the symplectic one-form density \cite{Crnkovic:1987tz,Lee:1990nz}, which is a one-form in the space of field variations and a $d$ form in spacetime, and $E_\phi$ is the equation of motion. Since we want to work with on-shell configurations, when we integrate this formula over the Euclidean asymptotically $AdS$ bulk $X$, the whole contribution will be a boundary term. In this way the only physical variations are those of the boundary condition $\delta \tilde{\lambda}$
. This gives
\beq
{\tilde \delta} S^{\rm on-shell}_{\rm grav}[\tilde \lambda ] 
= \int_{\partial X=S^d} \mathbf{\theta}(\tilde \lambda ,\delta \tilde \lambda).
\eeq
Now because of \eqref{eq:lambdatilde} and linearity of $\theta$ in the variation, if we do a strictly (anti)holomorphic variation, the integral localizes on the (upper)lower hemisphere
\beq
\delta_1 S = \int_{t_E<0} \mathbf{\theta}(\tilde \lambda ,\delta \tilde \lambda_1), \quad \delta_1^* S = \int_{t_E>0} \mathbf{\theta}(\tilde \lambda ,\delta \tilde \lambda_1).
\eeq
We then obtain the K\"ahler form using \eqref{eq:qftkahler}
and the fact that variations commute 
\bea
\Omega (\delta_1 \tilde \lambda,\delta_2 \tilde \lambda) &= 
i([\delta_1+\delta_1^*]\delta_2^*-[\delta_2+\delta_2^*]\delta_1^*) S^{\rm on-shell}_{\rm grav}[\tilde \lambda ] \\
&=
i\left[{\tilde \delta}_1\int_{t_E>0} \mathbf{\theta}(\tilde \lambda ,\delta \tilde \lambda_2)-{\tilde \delta}_2\int_{t_E>0} \mathbf{\theta}(\tilde \lambda ,\delta \tilde \lambda_1)\right].
\eea
Now we use the extrapolate dictionary, relating sources $\tilde \lambda$ to the boundary values of the dual bulk fields $\phi$. Note that $\phi$ can be the metric when we are sourcing the stress tensor. 
Using that the bulk symplectic 2-form density is 
\beq
\label{eq:bulksympl}
\omega_{\rm bulk}(\phi;\delta\phi_1,\delta\phi_2) = \delta_1 \theta(\phi,\delta_2\phi)-\delta_2 \theta(\phi,\delta_1\phi),
\eeq
we arrive at
\beq
\label{eq:boundsympl}
\Omega =i \int_{(\partial X)^+} \omega_{\rm bulk}=i \int_{t_E>0} dx(\delta \lambda_1^*\delta_2 \langle O \rangle-\delta \lambda_2^*\delta_1 \langle O \rangle) 
\eeq
where $(\partial X)^+$ denotes the $t_E>0$ part of the boundary of $X$. One could write an equivalent formula using only the lower part $(\partial X)^-$. \ In \eqref{eq:boundsympl}, using the usual dictionary we think of  $\delta \lambda, \delta \langle O \rangle$ (which are canonical conjugates) in terms of the asymptotic values of $\delta \phi$.
 
\subsection{Pushing into the bulk}

The expression derived above is integrated on part of the boundary of the Euclidean manifold, while the gravitational symplectic form should be integrated on a Cauchy slice. When the fields satisfy the equations of motion, the symplectic form is conserved $d\omega_{\rm bulk}(\phi,\delta \phi_1,\delta \phi_2)=0$ and can be pushed to other codimension $1$ surfaces. The relation between $\delta \phi$ and the variations of the sources is formally given by the Euclidean boundary to bulk propagator $G_E(Y|y)=\langle \Phi(Y) O(y) \rangle_{\tilde \lambda}$:
\bea
\label{eq:bulkvariations}
\delta \phi(Y) & =\delta \phi^{+}(Y)+\delta \phi^{-}(Y)
\eea
with
\bea
\label{eq:phipm}
\delta \phi^\pm(Y)=\int_{(\partial X)^\pm}dyG_E(Y|y)\delta  \tilde{\lambda}(y)
\eea
with $\tilde{\lambda}$ defined in \eqref{eq:lambdatilde}.
Here, $Y$ are bulk coordinates and we are using the standard dictionary which relates bulk classical configurations to boundary sources. 

\begin{figure}[h!]
\centering
\includegraphics[width=0.45\textwidth]{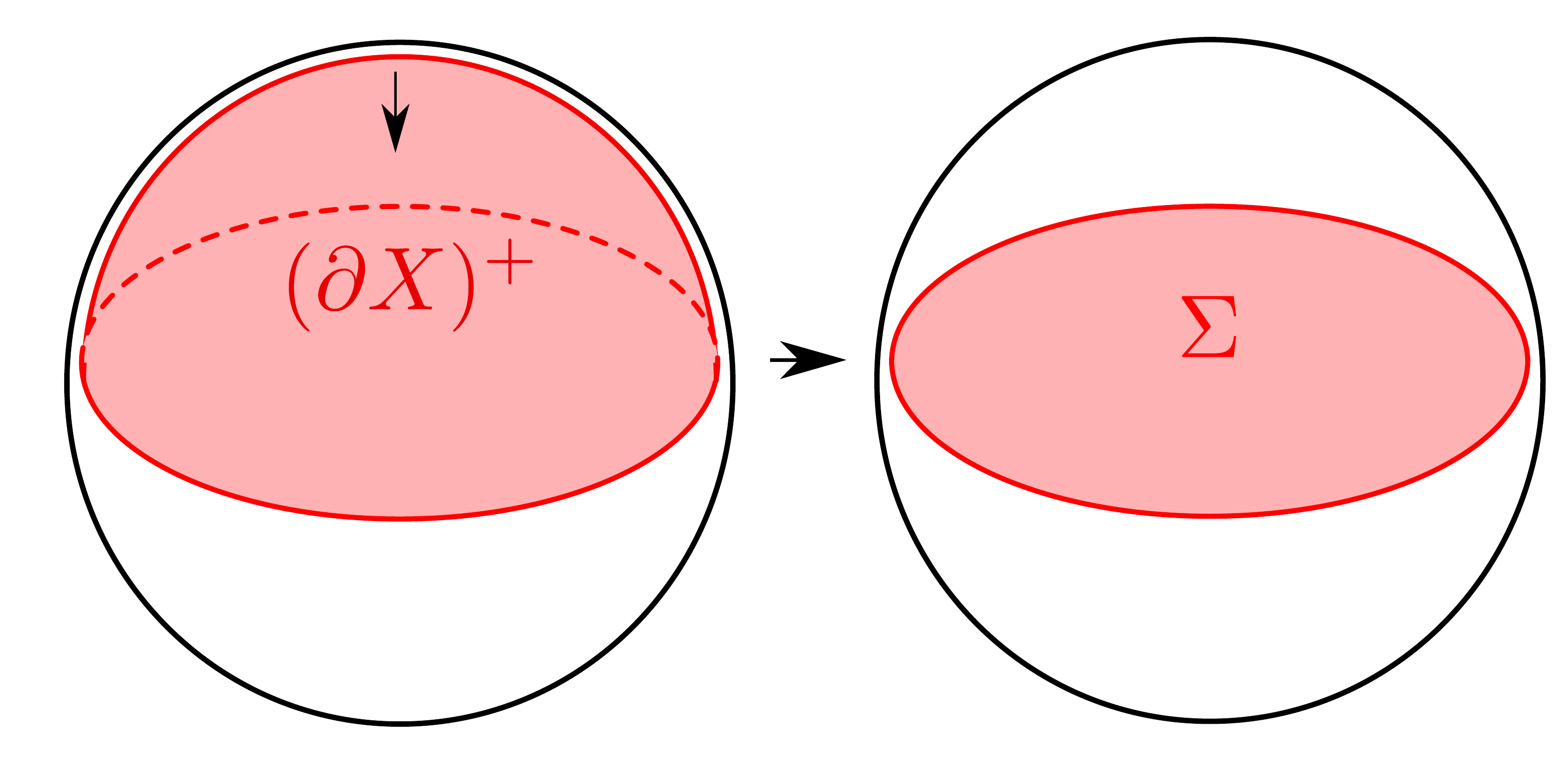}
\caption{Using conservation we can push the symplectic form from the Euclidean upper hemisphere $(\partial X)^+$ to an arbitrary slice $\Sigma$ anchored at $t_E=0$.}
\label{fig:1}
\end{figure}

We can now push the surface $(\partial X)^+$ in \eqref{eq:boundsympl} to the Euclidean bulk. When $ \partial X$ is a sphere, $(\partial X)^+$ is simply the northern hemisphere such that we can push it to any surface $\Sigma$ that is anchored at $t_E=0$, see Fig \ref{fig:1}.
 However, to relate it to the bulk symplectic structure, we need to integrate the symplectic flux on a Lorentzian initial value surface. Consider the case when the background $\lambda$ is real, the boundary condition $\tilde \lambda$ (and hence the bulk configuration) is $Z_2$ symmetric and there is a $Z_2$ invariant slice $\Sigma_0$ that is special. This slice continues nicely to Lorentzian since all fields are real there and their time derivatives are all zero because of the $Z_2$ symmetry. Therefore, 
we can pick variations in the complexified tangent space that correspond to real Lorentzian initial data. Reality of this data is precisely the requirement that $\delta \lambda^*$ is the complex conjugate of $\delta \lambda$. One can see this by looking at \eqref{eq:bulkvariations} and using that $Z_2$ symmetry of the slice implies $G_E(Y_0|y)=G_E(Y_0|y^T)$ and $\partial_n G_E(Y_0|y)=-\partial_n G_E(Y_0|y^T)$ (here $\partial_n$ is the normal derivative to $\Sigma_0$) for any $Y_0$ lying on $\Sigma_0$, from which it follows that 
\bea
\delta \varphi(Y_0) 
&=\text{Re} [\delta \phi^+(Y_0)] ,& \delta \pi(Y_0) = \text{Im} [\partial_n \delta \phi^+(Y_0)]. \label{phipi}
\eea
Here, $\pi$ are the conjugate momenta to the initial data $\varphi$, as defined through the bulk Lagrangian. Note that we use $\phi$ for time dependent field configurations and $\varphi$ for initial data.
In this way, the main result of the paper is:

    \bea
\label{eq:mainresult}
\Omega (\delta \tilde  \lambda_1,\delta \tilde \lambda_2) =  \int_{\Sigma_0} \omega_{\rm Lor}(\phi,\delta \phi_1,\delta \phi_2),
\eea
that is, the boundary symplectic form is given in the bulk by the symplectic form of the initial data in the Lorentzian $t=0$ time slice $\Sigma_0$ and the position and momenta are determined by $\delta \phi$ through \eqref{phipi}.

To create more general states, we need to consider complex $\lambda$. In this case, the boundary $t=0$ surface has $Z_2+{\cal C}$ symmetry, where ${\cal C}$ is complex conjugation. If we assume that this symmetry extends to the bulk such that it acts pointwise, we can consider surfaces $\Sigma$ which are fixed by this symmetry. By construction, one has $\text{Im}\varphi=0$ and $\text{Re}\pi=0$ on such surfaces. Similarly, the extrinsic curvature must be imaginary, while the induced metric must be real on the surface. These therefore correspond to real Lorentzian initial data. The Lorentzian variations \eqref{eq:bulkvariations} are real again if $\delta \lambda^*$ is the complex conjugate of $\delta \lambda$, provided we assume that $G_E$ is invariant under the action of $Z_2+{\cal C}$. Given that these $Z_2 + {\cal C}$ symmetric geometries are complex, there is in principle a family of such geometries, which correspond to different choices of bulk Lorentzian slices, since $\Sigma$ will not have vanishing extrinsic curvature tensor. From the requirement of having well defined Lorentzian data, it seems like the maximal volume slice should be preferred \cite{Witten:2018lgb}. We will analyze these situations in more detail in \cite{BLS}.

\subsection{Complex structure and K\"ahler metric}
From \eqref{eq:bulkvariations}, the action of the complex structure \eqref{eq:complexstructure} is nonlocal on the bulk variations. Regardless, it still has a nice interpretation, as hinted from calculating the norm of a variation in the K\"ahler metric \eqref{eq:kahlermetric}
\bea
g(\delta_1 \tilde \lambda,\delta_1 \tilde \lambda) &=\frac{1}{2}\Omega(\delta_1 \tilde \lambda,J(\delta_1 \tilde \lambda)) \\ &=\int_\Sigma \omega_{{\rm Lor}} \Big(\phi, \delta_1 \phi^-(Y), [\delta_1 \phi^-(Y^T)]^*\Big),
\eea
where in the last line we have assumed that the slice $\Sigma$ is $Z_2+{\cal C}$ symmetric, which implies that when $Y$ is close to $\Sigma$, we have $G_E^{*}(Y^T|y^T)=G_E(Y|y)$.
We used the action of $J$ on the sources \eqref{eq:complexstructure}
and we dropped the diagonal terms $\delta_1 \lambda \delta_1\lambda$ and $\delta_1 \lambda^* \delta_1\lambda^*$ since we know from the field theory expression \eqref{eq:qftkahler} that they must cancel. Notice that this expression, when continued to Lorentzian, is the Klein-Gordon norm for the projected variation $\delta_1 \phi^-$, defined in \eqref{eq:phipm}, which is sourced only from the lower hemisphere. By construction, the K\"ahler metric is positive definite, so we propose to interpret $\delta \phi^-=\frac{(1+J)}{2} \delta \phi$ as the generalization (in the absence of time translation symmetry) of the positive frequency part of the solution $\delta \phi$, which is important for defining the quantization of field theory in curved spacetime \cite{Wald:1995yp}. The relation between positive frequencies and negative Euclidean times is manifest for perturbations around empty AdS, where one can go to Fourier space \cite{Marolf:2017kvq}. Note that we do not have an explicit bulk alternative expression for the variation in the K\"ahler metric, other than projecting the field variation onto holomorphic sources,  but it is natural to expect this to be the subspace on which the Klein-Gordon norm is positive. 

\section{Special variations}

In this section, we will explore some particular examples of variations which have nice interpretations. We will consider the equation \eqref{eq:mainresult} directly in Lorentzian, for arbitrary states (not necessarily time reflection symmetric). 
\subsection{Time translations}
We can run a basic sanity check on \eqref{eq:mainresult} by recovering Wald's conserved charges \cite{Iyer:1994ys}. When the variation is a diffeomorphism, the symplectic form becomes a boundary term, which is interpreted as the boundary conserved energy. Let us take the variation of the bulk initial data to be a small translation along the normal vector field $\xi$ to our Cauchy slice, i.e. $\delta \phi= \mathcal{L}_\xi \phi$. This is a pure diffeomorphism, but because it is non-trivial at the boundary, it acts physically on the Hilbert space: it is a time derivative. We can regard this variation as a purely imaginary complex source, since $|\psi(t) \rangle=e^{-i H t}|\psi\rangle$.\footnote{This is a boundary diffeomorphism at positive times, which is equivalent to inserting the Hamiltonian: $i \int_{t>0} dy \nabla^{a} \xi^{b} T_{a b}=i \int_{t=0}  d\Sigma^b \xi^{a} T_{a b}=i H[\xi]$.} In this way, we have:
\begin{equation}
\label{eq:Hamiltondef}
    \Omega(\delta \tilde \lambda,{\cal{L}}_{\xi} \tilde \lambda)= (\delta_{\lambda}+\delta_{\lambda^*})\frac{\langle \lambda|H|\lambda \rangle}{\langle \lambda |\lambda \rangle} \equiv  \delta \langle H \rangle,
\end{equation}
since the time derivative gets opposite sign contributions when acting on the holomorphic/antiholomorphic sources. 
Comparing with \eqref{eq:mainresult} gives
\beq
\delta \langle H \rangle = \int_\Sigma \omega_{\rm Lor}(\phi,\delta \phi,\mathcal{L}_\xi \phi),
\eeq
which is just the usual covariant phase space definition of the Hamiltonian.


\subsection{Volume of an extremal slice}

In Einstein gravity, the symplectic form reads as
\bea
\Omega(\delta_1 \tilde{\lambda},\delta_2 \tilde{\lambda}) &= \int_{\Sigma} \big( \delta_1  h_{ab} \delta_2 p^{ab}-\delta_2  h_{ab} \delta_1 p^{ab}\big)\\
p_{ab}&=\sqrt{h}(K_{ab}-h_{ab} K),
\eea
where $h_{ab}$ is the induced metric on $\Sigma$ and 
$
K_{ab}=-\frac{1}{2}(\nabla_{a}n_b+\nabla_b n_a),
$
is its extrinsic curvature ($n_a$ is the outward pointing normal from $\Sigma$). Consider the Weyl-like transformation\footnote{A similar transformation arises as a diffeomorphism for spheres in flat space in  \cite{Jacobson:2015hqa}.}
\beq
\label{eq:weyl}
\delta_w h_{ab}=0, \quad \delta_w K_{ab}=\alpha h_{ab},
\eeq
where $\alpha$ is a small, dimensionful parameter. Denoting $\delta \tilde  \lambda_w$ the corresponding tangent space vector we have
\bea
\Omega(\delta_w \tilde  \lambda,\delta \tilde  \lambda)&= \alpha (d-1) \int_\Sigma \sqrt{h}h^{ab}\delta h_{ab}\\
&=2\alpha(d-1) \delta V,
\eea
where $V$ is the volume of the slice $\Sigma$. In order for \eqref{eq:weyl} to be an allowed deformation of initial data, the perturbed data must satisfy the momentum and Hamiltonian constraints 
\bea
\delta_w ( \nabla^j K_{jk}-\nabla_k K^j_j) = \alpha ( \nabla^j h_{jk}&-\nabla_k h^j_j) =0 \\
\delta_w(K_{ij}K^{ij}-(K)^2-R_{d}-d(d-1))&=2\alpha (d-1)K.
\eea
Note that in the first equation, $\delta_w \nabla_k=0$ since these derivatives depend only on $h_{ij}$ and $\delta_w h_{ij}=0$. The second equation is not satisfied automatically, it requires $K=K^i_i=0$, which is equivalent with the surface $\Sigma$ being extremal. Therefore, we can only access the variation of the volume of extremal slices. In order to have a complete CFT description of this volume, we need to understand the variation of the boundary background metric $(\delta_w \lambda_{ab}, \delta_w \lambda_{ab}^*)$ that induces \eqref{eq:weyl}. In the case where the geometry dual to the state $|\lambda\rangle$ is known, these are in principle obtainable from the standard dictionary \eqref{eq:bulkvariations}.
However, it is clearly desirable to have a more general understanding of these variations and why they are special. We will explore this and its relations with complexity \cite{Susskind:2014rva} in more detail in \cite{BLS}.

This boundary expression for the volume is purely Euclidean and it is different from what one would get using HKLL \cite{Hamilton:2006az} in Lorentzian: while the usual HKLL expression would be very complicated in the boundary -- it depends on the change of the boundary vev's through a very non-trivial kernel -- the present expression is simple in terms of the boundary sources and vevs. It seems unlikely that such a simplification could be obtained using Lorentzian techniques since it heavily relies on the use of the symplectic form.

\section{Conclusions}

In this work, we have studied the overlaps of nearby path integral states in holographic quantum field theories. We have found that in the field theory side, this gives rise to a K\"ahler form on the space of Euclidean sources which on the bulk side is dual to the gravitational symplectic form. This gives a precise duality between overlaps in the boundary and codimension one surfaces in the bulk. We have considered the case when the boundary has the topology of a ball, but it is clear that we can push the symplectic form for more complicated topologies, as long as the topology of the $t=0$ slice is the same as that of the boundary for positive Euclidean times (see \cite{Maloney:2015ina} for a set of coordinates which foliate geometries with arbitrary boundary topologies in $d=2$). Of course, since our formalism is completely covariant, it applies to any theory of gravity using Wald's formalism \cite{Iyer:1994ys}, so this matching is true for general theories of gravity.

There are many immediate questions that we have not touched upon. In an upcoming work \cite{BLS}, we will discuss some of these, such as the JKM ambiguities in the bulk symplectic flux \cite{Jacobson:1993vj}, bulk quantum corrections and a more detailed discussion of the dual description for the volume of the maximal slice. It would be interesting to investigate the relation of our work to the holographic renormalization group \cite{deBoer:1999tgo}, where the relevant Euclidean evolution does not keep the $t=0$ slice fixed.

\section*{Acknowledgements}
We are happy to thank Jan de Boer, Ted Jacobson, Alex Maloney, Onkar Parrikar, Tadashi Takayanagi and Herman Verlinde for useful conversations. We thank the Galileo Galilei Institute for Theoretical Physics for the hospitality and the INFN for partial support during the completion of this work. AB is supported by the NWO VENI grant 680-47-464 / 4114. AL and GS acknowledge support from the Simons Foundation through the It from
Qubit collaboration (385592, V. Balasubramanian). AL would also like to thank the David Rittenhouse Laboratory for hospitality during the development of this work. This work is supported by the Delta ITP consortium, a program of the Netherlands Organisation for Scientific Research (NWO) that is funded by the Dutch Ministry of Education, Culture and Science (OCW) and by the Belgian Federal Science Policy Office through the Interuniversity Attraction Pole P7/37, by FWO-Vlaanderen through projects G020714N and G044016N, by Vrije Universiteit Brussel through the Strategic Research Program ``High-Energy Physics'' and by the US DOE through Grant FG02-05ER-41367.

\bibliographystyle{utphys}
\bibliography{symplectic}
\end{document}